\documentclass[tighten]{aastex63}
\usepackage{here}
\usepackage{amsmath}
\usepackage{natbib}
\usepackage{enumitem}

\shorttitle{Time Evolution of Emission from an M-Class Flare}
\shortauthors{Nagasawa et al.}

\graphicspath{{./}{figures/}}

\begin{document}

\title{Study of Time Evolution of Thermal and Non-Thermal Emission 
from an M-Class Solar Flare}

\correspondingauthor{Shunsaku Nagasawa}
\email{shunsaku.nagasawa@ipmu.jp}

\author[0000-0001-6574-6784]{Shunsaku Nagasawa}
\affiliation{Department of Physics, University of Tokyo, 7-3-1 Hongo, Bunkyo, Tokyo 113-0033, Japan}
\affiliation{Kavli Institute for the Physics and Mathematics of the Universe (Kavli IPMU, WPI)\\
The University of Tokyo, 5-1-5 Kashiwanoha, Kashiwa, Chiba 277-8583, Japan}

\author[0000-0002-1021-0322]{Tomoko Kawate}
\affiliation{National Institute for Fusion Science, 322-6 Oroshi-cho, Toki, Gifu 509-5292 JAPAN}
\affiliation{The Graduate University for Advanced Studies (SOKENDAI), Shonan-Kokusai Village, Hayama, Miura, Kanagawa 240-0193 Japan}
\affiliation{National Astronomical Observatory of Japan, Mitaka, Tokyo 181-8588, Japan}

\author[0000-0002-6330-3944]{Noriyuki Narukage}
\affiliation{National Astronomical Observatory of Japan, Mitaka, Tokyo 181-8588, Japan}
\affiliation{The Graduate University for Advanced Studies (SOKENDAI), Shonan-Kokusai Village, Hayama, Miura, Kanagawa 240-0193 Japan}

\author[0000-0001-6305-3909]{Tadayuki Takahashi}
\affiliation{Kavli Institute for the Physics and Mathematics of the Universe (Kavli IPMU, WPI)\\
The University of Tokyo, 5-1-5 Kashiwanoha, Kashiwa, Chiba 277-8583, Japan}
\affiliation{Department of Physics, University of Tokyo, 7-3-1 Hongo, Bunkyo, Tokyo 113-0033, Japan}

\author[0000-0001-8702-8273]{Amir Caspi}
\affiliation{Southwest Research Institute, 1050 Walnut St Suite 300, Boulder, CO 80302}

\author[0000-0002-2308-6797]{Thomas N. Woods}
\affiliation{Laboratory for Atmospheric and Space Physics, University of Colorado at Boulder, 3665 Discovery Dr., Boulder, CO 80303}

\begin{abstract}
We conduct a wide-band X-ray spectral analysis in the energy range of 1.5--100~keV to study the time evolution of the M7.6 class flare of 2016 July 23, with the {\it Miniature X-ray Solar Spectrometer (MinXSS)} CubeSat and the {\it Reuven Ramaty High Energy Solar Spectroscopic Imager (RHESSI)} spacecraft. 
With the combination of {\it MinXSS} for soft X-rays and {\it RHESSI} for hard X-rays, a non-thermal component and three-temperature multi-thermal component -- ``cool" ($T \approx$ 3~MK), ``hot" ($T \approx$ 15~MK), and ``super-hot" ($T \approx$ 30~MK) -- were measured simultaneously.
In addition, we successfully obtained the spectral evolution of the multi-thermal and non-thermal components with a 10~s cadence, which corresponds to the Alfv\'en time scale in the solar corona. We find that the emission measures of the cool and hot thermal components are drastically increasing more than hundreds of times and the super-hot thermal component is gradually appearing after the peak of the non-thermal emission.
We also study the microwave spectra obtained by the {\it Nobeyama Radio Polarimeters (NoRP)}, and we find that there is continuous gyro-synchrotron emission from mildly relativistic non-thermal electrons.
In addition, we conducted a differential emission measure (DEM) analysis by using Atmospheric Imaging Assembly (AIA) onboard the {\it Solar Dynamics Observatory (SDO)} and determine that the DEM of cool plasma increases within the flaring loop.
We find that the cool and hot plasma components are associated with chromospheric evaporation. The super-hot plasma component could be explained by the thermalization of the non-thermal electrons trapped in the flaring loop.

\end{abstract}

\section{Introduction} \label{sec:intro}
Solar flares are powerful explosions, releasing coronal magnetic energy up to $\sim$10$^{33}$ ergs on short time scales (100--1000~s) and efficiently accelerating electrons up to several hundreds of MeV and ions to tens of GeV \citep{holman2011implications}. 
It has been established that magnetic reconnection plays an important role during solar flares \citep{shibata2011review}. 
Magnetic reconnection is a process in which oppositely oriented components of the magnetic field annihilate, the magnetic field reconfigures to a lower-energy state, and the liberated free energy of the magnetic field in the plasma is efficiently converted into particle kinetic energy through acceleration  and plasma heating \citep[see][for a review]{hesse2020magnetic}.
However, the total amount of magnetic energy released by magnetic reconnection and the proportion of distributed energy to the non-thermal particles and plasma heating remains poorly understood.

During solar flares, the energy released through magnetic reconnection is converted into other forms through processes such as heating of coronal plasma, bulk flows within coronal mass ejections, and particle acceleration \citep[see][for a review]{benz2017flare}. 
In addition, accelerated particles secondarily contribute to plasma heating through their collisions with the ambient plasma. In this way, the heating, cooling, and particle acceleration processes should be closely related to the magnetic reconnection and correlated to each other. Therefore, to resolve such a complicated energy conversion system, it is crucial to separate and follow the time evolution of ``thermal emission" from heated plasma and ``non-thermal emission" from accelerated electrons \citep{shibata1996new,holman2011implications} as a first step.

The multi-thermal structure of flares has been studied in the extreme ultraviolet (EUV: $E \approx 0.01$--0.2~keV) band using emission lines from multiply-ionized Fe, e.g., with the EUV Variability Experiment \citep[EVE;][]{woods2010extreme} on the {\it Solar Dynamics Observatory (SDO)}. \citet{warren2013observations} analyzed differential emission measure (DEM) distributions using \ion{Fe}{15}--\ion{Fe}{24} lines observed by {\it SDO}/EVE and showed that the isothermal approximation is not an appropriate representation of the thermal structure. However, EUV observations alone have poor sensitivity to thermal emission from plasmas hotter than $\sim$15--20~MK \citep{winebarger12}, particularly for ``super-hot" temperatures \citep[$T \gtrsim 30$~MK, e.g.,][]{caspi2010rhessi}, and are also not sensitive to non-thermal emission.
Moreover, since the timescale to reach ionization equilibrium for line emission may be longer than the timescales of relevant dynamic processes, spectral analysis using continuum emission -- which is only weakly sensitive to ionization state -- is more suitable to study the detailed time evolution of these processes.

The time evolution and relationships of thermal and non-thermal emission in flares have been studied using continuum emission -- bremsstrahlung (free-free) and radiative recombination (free-bound) -- from the heated plasma and accelerated electrons, e.g., observed by the {\it Reuven Ramaty High Energy Solar Spectroscopic Imager} \citep[{\it RHESSI};][]{Lin2002} spacecraft in hard X-rays (HXR: $E \gtrsim 10$~keV).
However, {\it RHESSI} is limited in its sensitivity to the soft X-ray (SXR: $E \lesssim 10$~keV) band, which is generally dominated by thermal emission, particularly from plasma with temperatures of $\sim$5--20~MK. During intense flares, attenuators are often inserted in front of the detectors to avoid pulse pile-up and preserve sensitivity to HXRs, especially during the impulsive phase of flares \citep{smith2002rhessi}. The absorption by attenuators makes it difficult to determine the exact shape of the low-energy spectrum and significantly increases the uncertainty of models fit to that region of the spectrum.
Consequently, it is difficult to resolve a multi-temperature structure, and the temperature and the emission measure of the cooler portion of the thermal emission often have to be predicated on the assumption of isothermal based on the {\it Geostationary Operational Environmental Satellite (GOES)} 2-channel X-Ray Sensor (XRS) SXR photometer fluxes \citep{white2005updated}.

Therefore, solar SXR spectral observations with high energy resolution ($\lesssim$1~keV FWHM) and high time resolution ($<$10~s,  comparable to the Alfv\'en time scale) are required for a precise characterization of such a multi-thermal structure and its relationship to non-thermal emission. 
However, most prior SXR observations have been carried out either with high spectral resolution only in narrow bandpasses to track specific ionization lines (e.g., using the Bragg Crystal Spectrometer (BCS) on {\it Yohkoh}) or through measurements of spectrally integrated fluxes over a large bandpass (e.g., using the XRS on {\it GOES}).

The {\it Miniature X-ray Solar Spectrometer} ({\it MinXSS}) CubeSat is the first mission that routinely archived solar flare spectral observations with high energy resolution ($\sim$0.15~keV FWHM) and high time resolution (10 s time cadence) in the SXR band \citep{2016JSpRo..53..328M,moore2018instruments}.
Utilizing {\it MinXSS} data, \citet{woods2017new} conducted a  spectral analysis for an M5.0 flare that occurred on 2016 July 23 that peaked at 02:11~UT. The {\it MinXSS} spectra obtained during the flare are generally well described by a two-temperature model with a cool and a hot component. However, because {\it MinXSS} has little sensitivity to HXR emission, studying the super-hot and non-thermal components at the same time requires analyzing {\it RHESSI} HXR spectra and {\it MinXSS} SXR spectra simultaneously.

In this paper, we conduct a wide-band X-ray spectral analysis using combined {\it MinXSS} SXR and {\it RHESSI} HXR data for understanding the thermal and non-thermal emissions in a solar flare.
In Section~\ref{sec:Observations}, we summarize the observations of the target flare we analyzed. In Section~\ref{sec:Analysis}, we introduce the data preparation to realize the wide-band X-ray spectral analysis and show the results. We also compare these results with microwave and EUV observations.
Based on these results, we discuss the origins of thermal and non-thermal emission in Section~\ref{sec:Discussion}, and summarize our conclusions in Section~\ref{sec:Conclusion}.

\section{Observations}\label{sec:Observations}
\begin{figure*}[htb]
\begin{center}
\includegraphics[width=1.0\hsize]{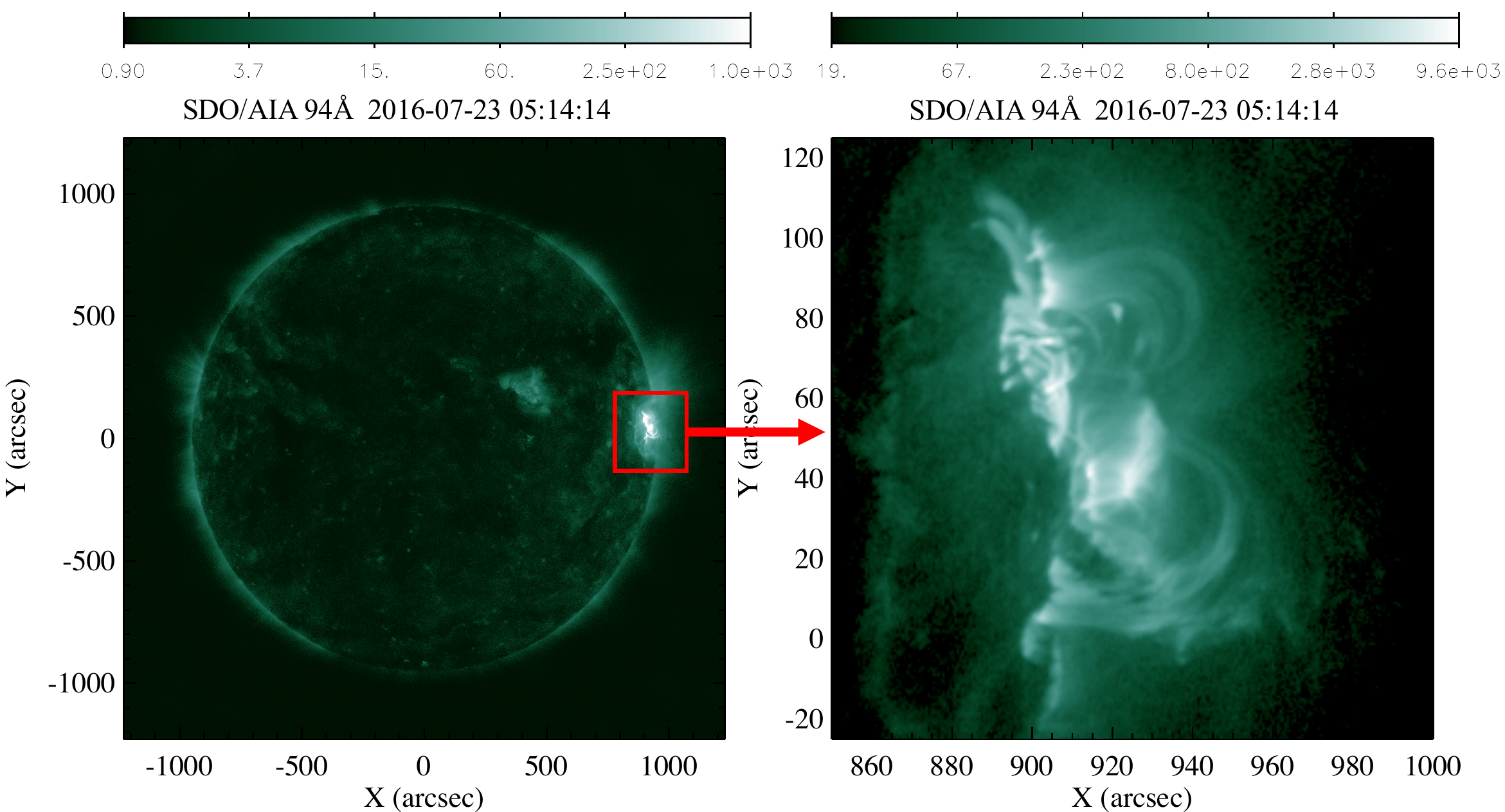}
\caption{{\it SDO}/AIA 94~{\AA} images of the GOES M7.6 flare at 2016-07-23 05:14:14~UT. 
The flare onset occurred around 05:00~UT in NOAA active region 12567 in the northern hemisphere and near the west limb (N05W73).}
\label{aia_fullsun}
\end{center}
\end{figure*}

We analyzed the GOES M7.6 class flare, which occurred starting around 05:00~UT on 2016 July 23. The flare is located in NOAA active region 12567 in the northern hemisphere and near the west limb (N05W73).
Figure~\ref{aia_fullsun} shows the 94~{\AA} EUV images of the flare during the impulsive phase taken by the Atmospheric Imaging Assembly \citep[AIA;][]{aia2011} onboard {\it SDO}. AIA takes full-solar images in three UV filters and seven EUV filters with 1.5{\arcsec} resolution and a cadence of 12~s. There are no other flares occurring on the Sun at this time. 

This flare is the most intense event during the one-year observation lifetime of {\it MinXSS}, from May 2016 to May 2017. {\it MinXSS} made spectral observations of the entire Sun (spatially integrated) with moderately high energy resolution ($\sim$0.15~keV FWHM) with a time cadence of 10~s in the SXR band of 0.8--12~keV \citep{moore2018instruments}. 
This flare was also observed by {\it RHESSI}, which provides spectral observations with $\sim$1~keV (FWHM) resolution and rotational modulation collimator imaging with angular resolution down to $\sim$2{\arcsec} \citep{Lin2002}.
The flare was observed through the thin (A1) or thick+thin (A3) attenuators during the impulsive phase of the flare to reduce the intense SXR flux.
There was no flare observation data for {\it MinXSS} after 05:21~UT and for {\it RHESSI} before 05:04~UT because of the spacecraft ``eclipse" time (when the satellite is in the shadow of the Earth), but there is simultaneous data from both instruments in the 05:04--05:21~UT period.
The XRS on {\it GOES} continuously measures solar SXR fluxes in two broad energy bands (XRS-A: 0.5--4.0~{\AA}~and XRS-B: 1.0--8.0~{\AA}) with a time cadence of $\sim$2~s \citep{Garcia1994}.
Microwave emission from the flare was observed by the {\it NobeyamaRadio Polarimeters} \citep[{\it NoRP};][]{nobeyama1985}, which measured total fluxes from the entire Sun at 1, 2, 3.75, 9.4, 17, 35, and 80~GHz with a time cadence of 0.1~s.

\begin{figure*}[htb]
\begin{center}
\includegraphics[width=1.0\hsize]{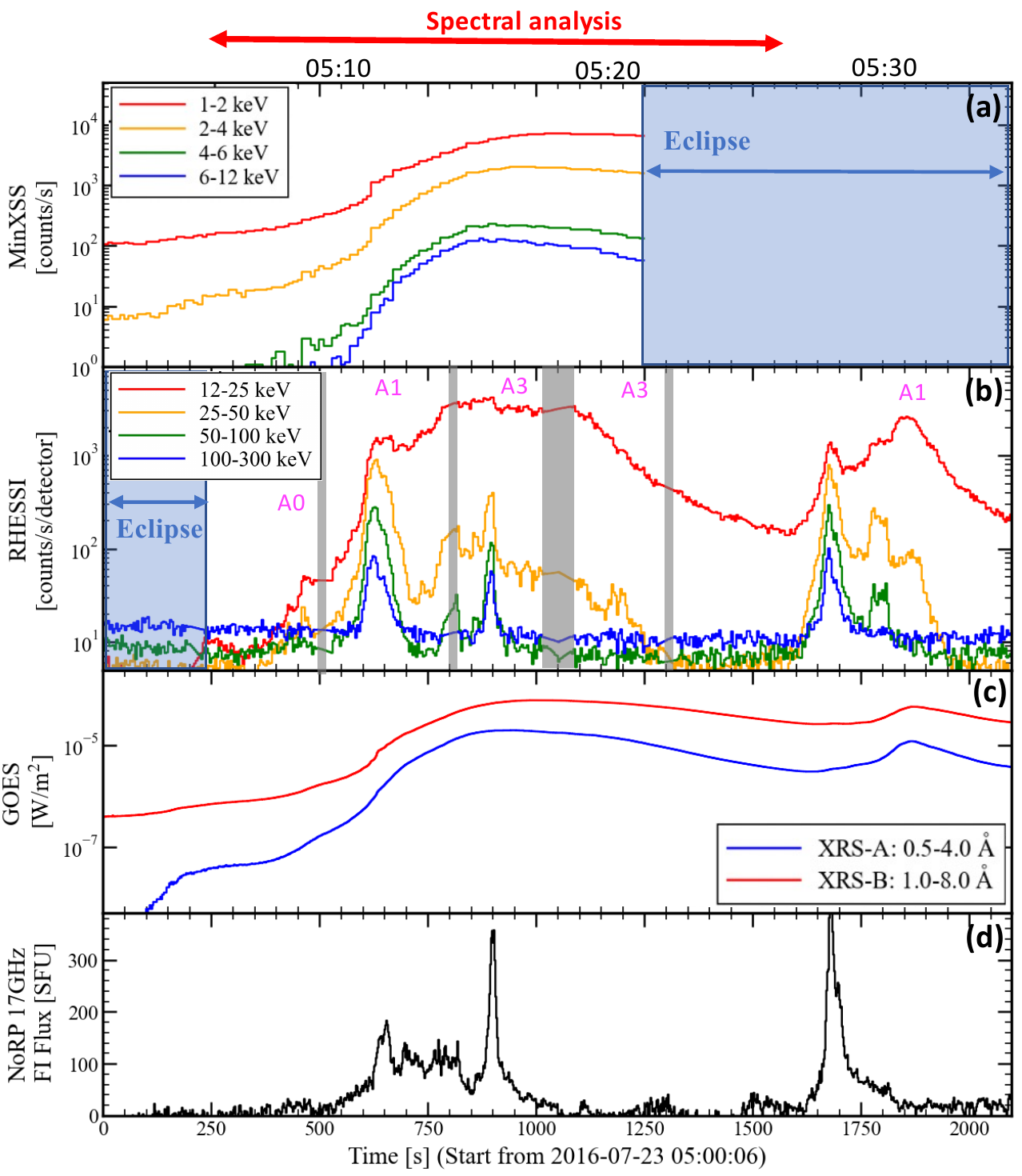}
\caption{Multi-wavelength lightcurves of the 2016 July 23 M7.6 solar flare. (a) Soft X-ray {\it MinXSS} count rates in 10~s time bins and several energy bands, as marked. (b) Hard-X-ray {\it RHESSI} corrected-count rate in 4~s time bins and several energy bands, as marked. The A0, A1, and A3 labels represent the attenuator state \citep{smith2002rhessi}, the effects of which have been deconvolved from the lightcurves. (c) Soft-X-ray {\it GOES} fluxes in 2~s time bins for the short (XRS-A: 0.5--4.0~{\AA}) and long (XRS-B: 1.0--8.0~{\AA}) wavelength channels.  (d) Microwave {\it NoRP} flux $I(R+L)$ at 17~GHz with the average pre-flare flux subtracted. The blue-highlighted times for {\it MinXSS} and {\it RHESSI} indicate spacecraft ``eclipse" (when the satellite is in the Earth's shadow), and the gray-highlighted time intervals for {\it RHESSI} indicate switching between attenuate states, which are excluded from this study as the detector responses are not well defined during these periods.}
\label{total_lc}
\end{center}
\end{figure*}

Figure~\ref{total_lc} shows the temporal evolution of the SXR fluxes obtained by {\it MinXSS} and {\it GOES}, the HXR flux measured by {\it RHESSI}, and the microwave intensity at 17~GHz observed by {\it NoRP}. 
The SXR flux gradually rises from the start of the flare (05:00:06~UT) and the {\it GOES} XRS-A (0.5--4.0~{\AA}) flux  reached its peak around 05:14~UT.
There are strong peaks around 05:12~UT and 05:15~UT in HXRs. Continuous microwave emission is also observed during the impulsive phase. After that, the SXR and HXR fluxes gradually decrease and then spike again around 05:28~UT.

\begin{figure}[htb]
\begin{center}
\includegraphics[width=0.8\hsize]{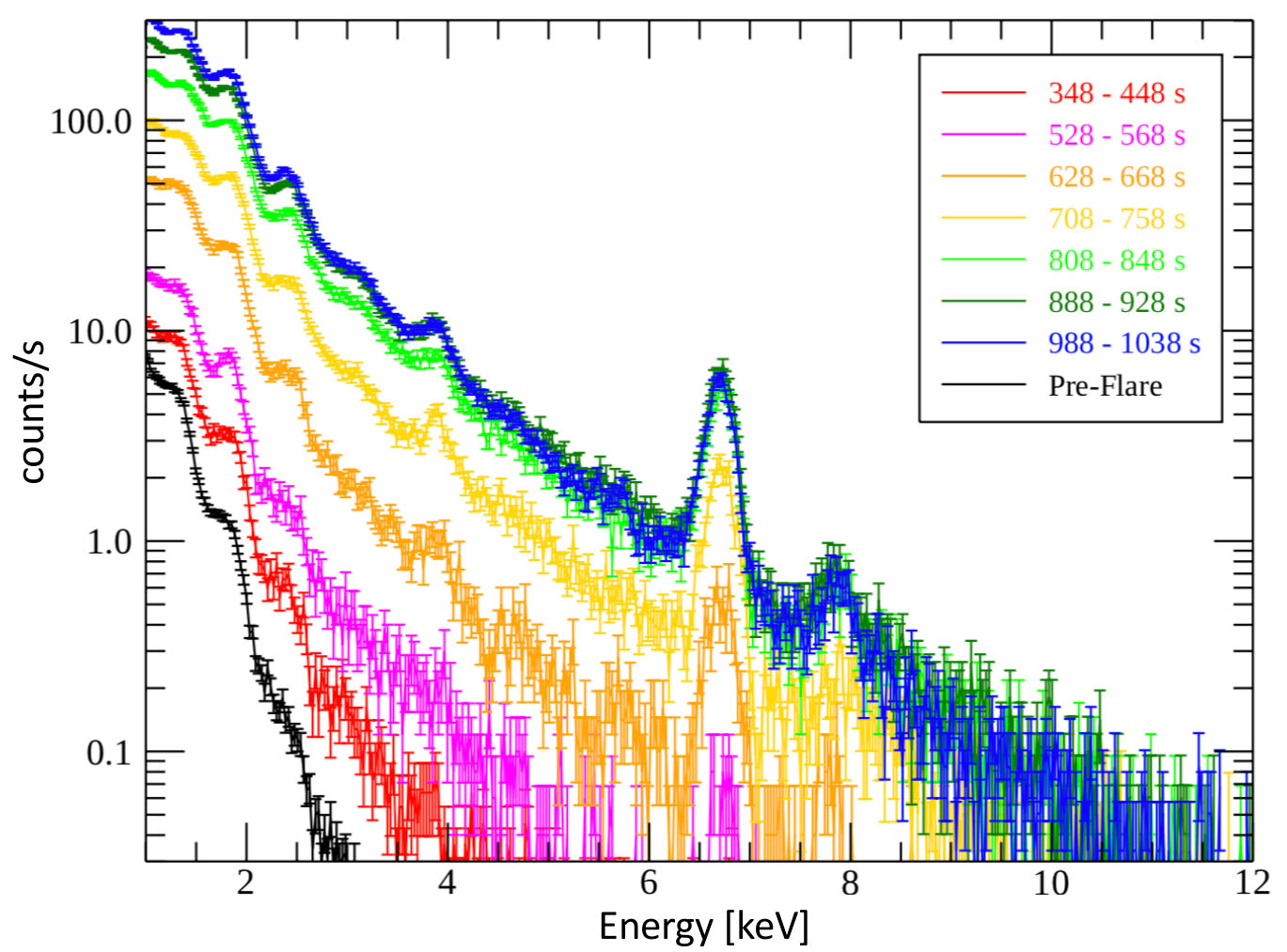}
\caption{Spectral evolution of the 2016 July 23 M7.6 solar flare observed by {\it MinXSS}. The time labels indicate the elapsed time from 2016-07-23 05:00:06~UT. The pre-flare spectrum before the start of the flare (04:45:54--04:57:34~UT) is also shown.}
\label{minxss_spect}
\end{center}
\end{figure}

Figure~\ref{minxss_spect} shows the spectral evolution in the SXR band observed by {\it MinXSS} during the flare. The detailed spectra in terms of energy and time evolution in the SXR band are obtained, and line emission from \ion{Ca}{19} (3.9 keV), \ion{Fe}{25} (6.7 keV, 7.8 keV), and \ion{Ni}{27} (7.8 keV) is clearly observed.
Therefore, by conducting spectral analysis and separately characterizing the thermal and non-thermal emission, we can follow the evolution of the temperature and emission measures of the thermal components, and the power-law index of the non-thermal emission.
This information will help to understand the mechanisms of particle acceleration and the heating and cooling processes associated with solar flares.

\section{Analysis}\label{sec:Analysis}
\subsection{Spectral Fitting using {\it MinXSS} and {\it RHESSI}}
We performed spectral fitting to {\it MinXSS} and {\it RHESSI} data simultaneously in the energy range of 1.5--100~keV. In this study, we utilize XSPEC \citep[version 12.11.0;][]{arnaud1996xspec}, the standard spectrum analysis tool in the field of high-energy astronomy (see Appendix \ref{app_xspec} for detail). The Object SPectral EXecutive (OSPEX)\footnote{\url{http://hesperia.gsfc.nasa.gov/ssw/packages/spex/doc/ospex_explanation.html}} software package in the SolarSoftWare (SSW)\footnote{\url{https://www.lmsal.com/solarsoft/ssw_install.html}} IDL suite is often used for X-ray spectrum analysis in solar physics. However, at this time, OSPEX cannot do simultaneous joint fitting of data from more than one instrument, and thus we use XSPEC which does allow such analysis. We note that there is no flare observation data for {\it MinXSS} after 05:20:54~UT because of spacecraft ``eclipse," so spectral analysis after this time is performed using only {\it RHESSI} data.

We use the forward modeling method in XSPEC, where an incident photon model spectrum is assumed and convolved with the instrument response to obtain a modeled count spectrum, which is then compared with the observed spectrum using the $\chi^2$-statistic to assess goodness of fit. The model parameters are then adjusted and the fit procedure is iterated until a minimum $\chi^2$ value is achieved. The statistical error of each channel count value is considered in the $\chi^2$ calculation, and the systematic error term in XSPEC is set to 0. The fit model components are the APEC isothermal emission model ({\tt vapec}), and a broken power-law ({\tt bknpower}) for non-thermal emission. {\tt vapec} models thermal emission from an optically-thin hot plasma and is calculated based on AtomDB \citep{foster2012updated}. The main parameters are a plasma temperature $T$ and an emission measure $EM$. Abundance ratios (atomic number ratios of each element relative to hydrogen) for He, C, N, O, Ne, Mg, Al, Si, S, Ar, Ca, Fe, and Ni can also be set or fit. In this study, we use abundances based on \citet{schmelz2012composition}.
{\tt bknpower} models power-law non-thermal emission with parameters including a break energy $E_{break}$, power-law photon indices of $\gamma_1$ for energies below $E_{break}$ and $\gamma_2$ for higher energies, and normalization $K$:
\begin{equation}
    A(E) = 
    \begin{cases}
    KE^{-\gamma_1}  & (E \leq E_{break}) \\
    KE_{break}^{-(\gamma_1-\gamma_2)}(E/1~\mathrm{keV})^{-\gamma_2} & (E \geq E_{break}) 
    \end{cases}
\end{equation}
Since thermal emission dominates at lower energies, the break energy $E_{break}$ is used to model the effective low-energy cutoff of the non-thermal emission. The spectral index below the break energy, $\gamma_1$, is held fixed at 2, and the other parameters are fitted.

The observed spectra are spatially integrated and therefore contain a background component. It is necessary to subtract this background to isolate the flare emission for spectral analysis. To subtract Non-Solar X-ray Background (NXB), which is mainly caused by bremsstrahlung from cosmic rays and charged particles interacting in the spacecraft, the NXB is evaluated by using spectral data during spacecraft ``eclipse." The NXB of {\it MinXSS} is negligible, but is significant for {\it RHESSI} and we thus time-average the spectra during the eclipse time of 04:54--05:02~UT and subtract this from the flare observations. The solar emission before the flare is treated as a ``pre-flare" background. This emission can be interpreted as X-rays emitted from the entire surface of the Sun other than the target solar flare. To isolate the flare emission, a time-average spectrum before the start of flare, integrated over 04:45:54--04:57:34~UT, is fixed as a ``pre-flare component" and incorporated into the model (added to the model flare emission) when conducting spectral analysis. While such background components can often vary in time for other flares, our assumption of a constant summed background is supported by the {\it RHESSI} lightcurves during eclipse and post-flare intervals (e.g., Figure~\ref{total_lc}).

\begin{figure}[htb]
\begin{center}
\includegraphics[width=1.0\hsize]{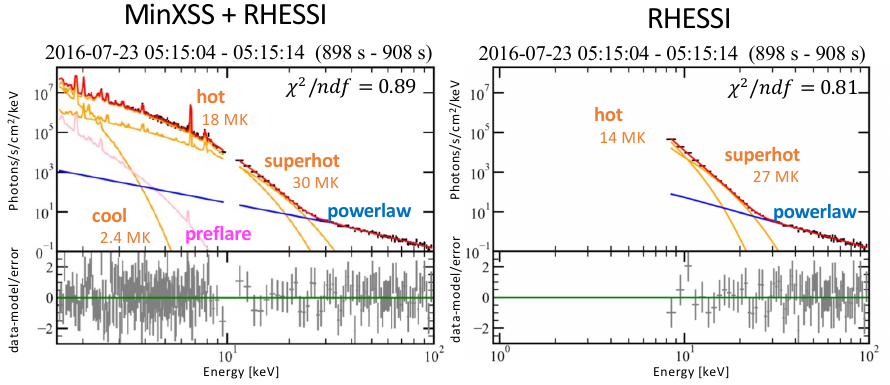}
\caption{Spectral fitting for the 2016 July 23 M7.6 solar flare during 05:15:04--05:15:14~UT, using {\it MinXSS} and {\it RHESSI} combined spectra (left) and with only the {\it RHESSI} spectrum (right). The addition of {\it MinXSS} SXR observations enable the multi-thermal structure to be clearly resolved.}
\label{minxss_vs_rhessi}
\end{center}
\end{figure}

Figure~\ref{minxss_vs_rhessi} shows the results of spectral fitting during the peak of the flare (05:15:04-05:15:14~UT) using both ``{\it MinXSS} and {\it RHESSI}" and ``only {\it RHESSI}" spectra. With {\it RHESSI} alone, the non-thermal power-law and the thermal emission from a ``hot" ($T \approx 15$~MK) and a ``super-hot" plasma ($T \approx 30$~MK) are detected. However, these two thermal components are poorly constrained and can ``trade off" with each other because of lack of sensitivity at lower energies, especially when the flare is in the impulsive phase and {\it RHESSI} is in attenuator state A3.
In contrast, with the addition of MinXSS spectra in the SXR band, the multi-thermal structure is resolved. Three thermal components -- a ``cool" plasma ($T \approx 3$~MK), a ``hot" plasma, and a ``super-hot" plasma -- and non-thermal power-law component are detected and constrained at the same time by simultaneously fitting ``{\it MinXSS} and {\it RHESSI}" spectra.

\begin{figure*}[p]
\begin{center}
\includegraphics[width=1.0\hsize]{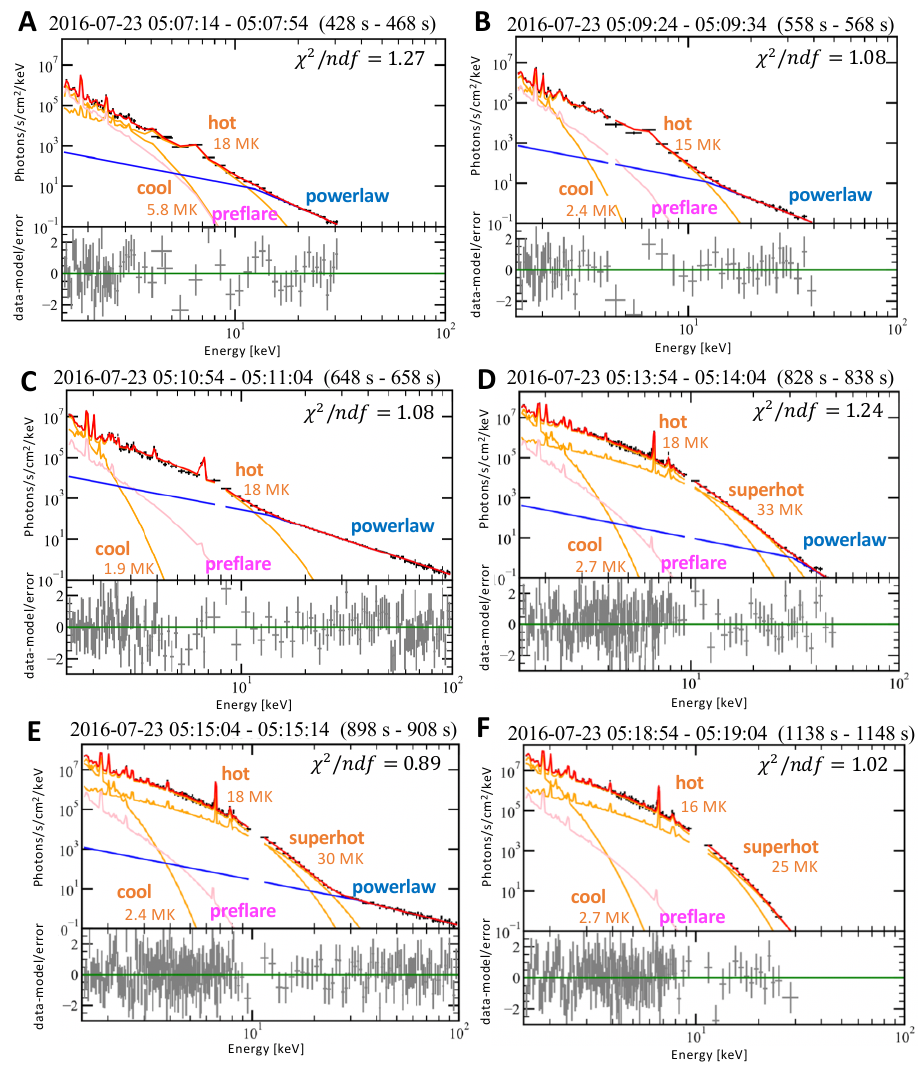}
\caption{Spectral evolution of the 2016 July 23 M7.6 solar flare using combined {\it MinXSS} SXR and {\it RHESSI} HXR spectra. The A--F labels correspond to the time intervals marked in Figure~\ref{total_para}. The pink spectrum represents the (fixed) pre-flare background. The blue curve represents the non-thermal emission (fit as a broken power law). The orange curves represents the thermal emission, and three temperatures are fit within the model: cool ($\sim$3~MK) and hot ($\sim$17~MK) components observed throughout the flare, and a super-hot ($\sim$30~MK) components starting around 05:11~UT. The red curve represents the sum of all model components: the pre-flare, thermal and non-thermal components. Animations of the spectral evolution for all time intervals are available in the online journal.}
\label{spect_long2}
\end{center}
\end{figure*}

Figure~\ref{spect_long2} shows the results of spectral fitting for each time interval using {\it MinXSS} and {\it RHESSI}, and the time evolution of the parameters of each of the thermal and non-thermal components are summarized in Figure~\ref{total_para}. The behavior of thermal and non-thermal emission in each flare phase is summarized in Table~\ref{flare_evol}. 
The comparison of the multi-temperature fitting model is shown in Appendix~\ref{app_spect}, for complete details. 

Even as early as the pre-impulsive phase of the flare (spectrum~A, B), isothermal emission alone is not sufficient to explain the observed spectrum. A cool thermal component ($T \approx 6$~MK) is also observed in addition to the hot thermal emission ($T \approx 18$~MK), which is inferred from the observed GOES fluxes, and a non-thermal component is also required at higher energies. In the first impulsive phase, as the HXR flux peaks and the non-thermal emission becomes harder, with $\gamma_2 \approx 2.8$ (spectrum~C), the emission measures of the hot and cool thermal emission increase drastically by more than hundreds of times. In addition, the temperature of cool plasma appears to decrease slightly to $T \approx 3$ MK. 
During the interval of the two impulsive phases of the flare, the non-thermal emission softens (higher $\gamma_2$), and the super-hot thermal emission ($T \approx 30$~MK) gradually appears (spectrum~D).
In the second impulsive phase, the spectral index $\gamma_2$ hardens as the HXR flux rises and then softens again as the HXR flux decreases. This Soft-Hard-Soft (SHS) behavior of non-thermal emission has been reported in other solar flares \citep[e.g.,][]{benz1977spectral,kosugi1988energetic}. The HXR flux then peaks again, and the non-thermal emission becomes the hardest, with $\gamma_2 \approx 2.6$ (spectrum~E). 
In the decay phase of the flare, the non-thermal emission fades and the temperatures of the hot and super-hot thermal emissions gradually cool (spectrum~F). 

\renewcommand{\baselinestretch}{1.1}

\begin{table}[hbtp]
  \caption{Time evolution of thermal and non-thermal emission of the GOES M7.6 Class flare}
  \label{flare_evol}
  \centering
  \begin{tabular}{p{1.5cm}|p{5.4cm}|p{4.7cm}|p{3.5cm}|p{1.3cm}}
     &Thermal Emission & \multicolumn{2}{l}{Non-Thermal Emission}  & \\
     \hline
    & Soft X-ray & Hard X-ray &  Microwave (17 GHz) & Spectrum\\
    \hline
    Pre-impulsive phase&
    Cool ($\sim$6 MK) and Hot ($\sim$18 MK) plasma components are detected.
    &
    The non-thermal power-law component is detected. 
    & The flux is not detected. &A, B\\
    \hline
    Impulsive phase I&
    The emission measures drastically increase ($>\times 100$).
    &
    The flux peaks and the power-law index becomes hard, $\gamma_2 \approx 2.8$.
    & The flux gradually increases and peaks.&C
    \\ \cline{1-5} 
    Interval phase&
    Super-hot ($\sim$30~MK) plasma component gradually appears.
    & Only small sub-peaks are detected.
    &
    The flux is continuously emitted.&D \\ \cline{1-5}
    Impulsive phase II& The emission measures continuously increase.&
    The flux peaks again, and becomes the hardest, $\gamma_2 \approx 2.6$.
     & The flux strongly peaks. &E \\
    \hline
    Decay phase& The temperatures of the hot and super-hot plasma components gradually cool and the emission measures also decrease. & The non-thermal power-law component fades. &
    The flux is not detected.&F\\
    \hline
  \end{tabular}
\end{table}

\renewcommand{\baselinestretch}{1}

\subsection{Comparison of fitting results with GOES flux}
In order to check the consistency of our fitting results using {\it MinXSS} and {\it RHESSI} with measured {\it GOES} fluxes, we estimated the X-ray fluxes that would be expected to be observed by {\it GOES} based on the spectral fits.  Then, we compared these estimated fluxes with those actually observed by {\it GOES}.

\begin{figure*}[p]
\begin{center}
\includegraphics[width=1.0\hsize]{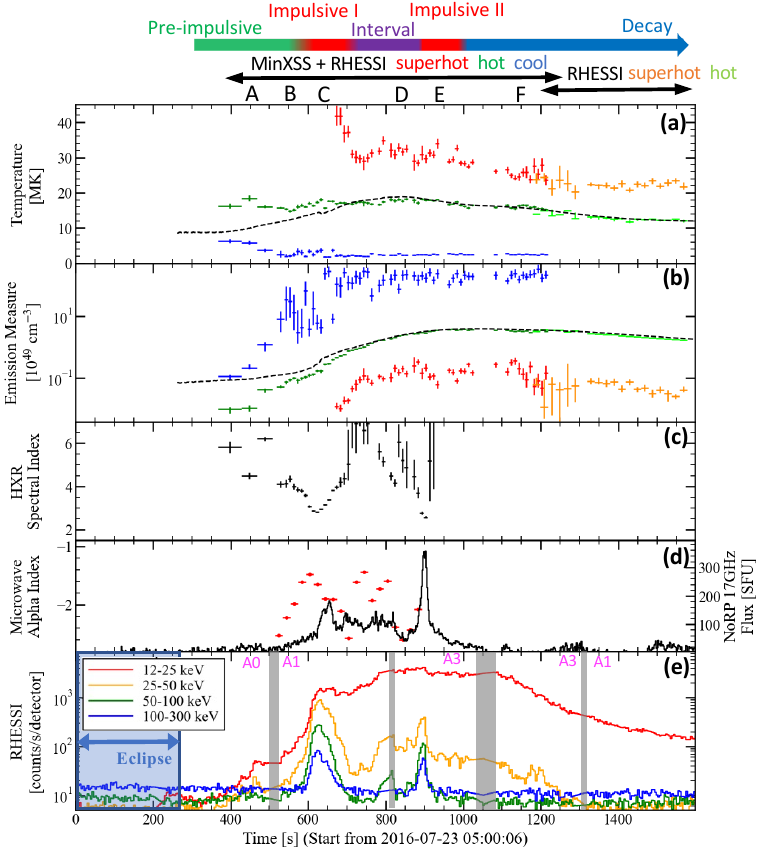}
\caption{Temporal evolution of thermal and non-thermal emission from spectral analysis. (a) Temperature and (b) emission measure of each fit thermal component. The black dotted curve represents the isothermal temperature and emission measure calculated from {\it GOES} two-channel fluxes \citep{white2005updated}. (c) HXR spectral index of non-thermal component above the break energy. (d) Spectral index of the NoRP microwave spectra above the turnover frequency (red), and NoRP 17~GHz flux (black). (e) Hard-X-ray {\it RHESSI} corrected-count rate, for reference. Error bars in (a)--(c) are 1$\sigma$, determined from the fit routine.}
\label{total_para}
\end{center}
\end{figure*}

First, the incident photon flux in units of $\mathrm{[W~m^{-2}]}$ is calculated based on the fit parameters for each time interval in Figure~\ref{total_para} (e.g., see the red spectral curves in Figure~\ref{spect_long2}, which are then converted to $\mathrm{W~m^{-2}}$). Then, the incident photon flux in each energy is converted to the current $I(E)~\mathrm{[A]}$ in the {\it GOES} ionization chamber by folding it through
the wavelength-dependent response of the {\it GOES} \citep[the transfer function, see Tables 6 and 7 of][]{goesres}.
The current at each energy $I(E)~\mathrm{[A]}$ is summed for all energies to obtain the total current, $I_{total}~\mathrm{[A]}$. 
Then, we divide the total current by the scalar flux conversion factor $C~\mathrm{[A~(W~m^{-2})^{-1}]}$ \citep[XRS-A: $1.342 \times 10^{-5}$, XRS-B: $5.703 \times 10^{-6}$; this includes the ``SWPC scaling factor," see Table 5 of][]{goesres} to estimate the expected {\it GOES} flux $\mathrm{[W~m^{-2}]}$.

\begin{figure}[htb]
\begin{center}
\includegraphics[width=1.0\hsize]{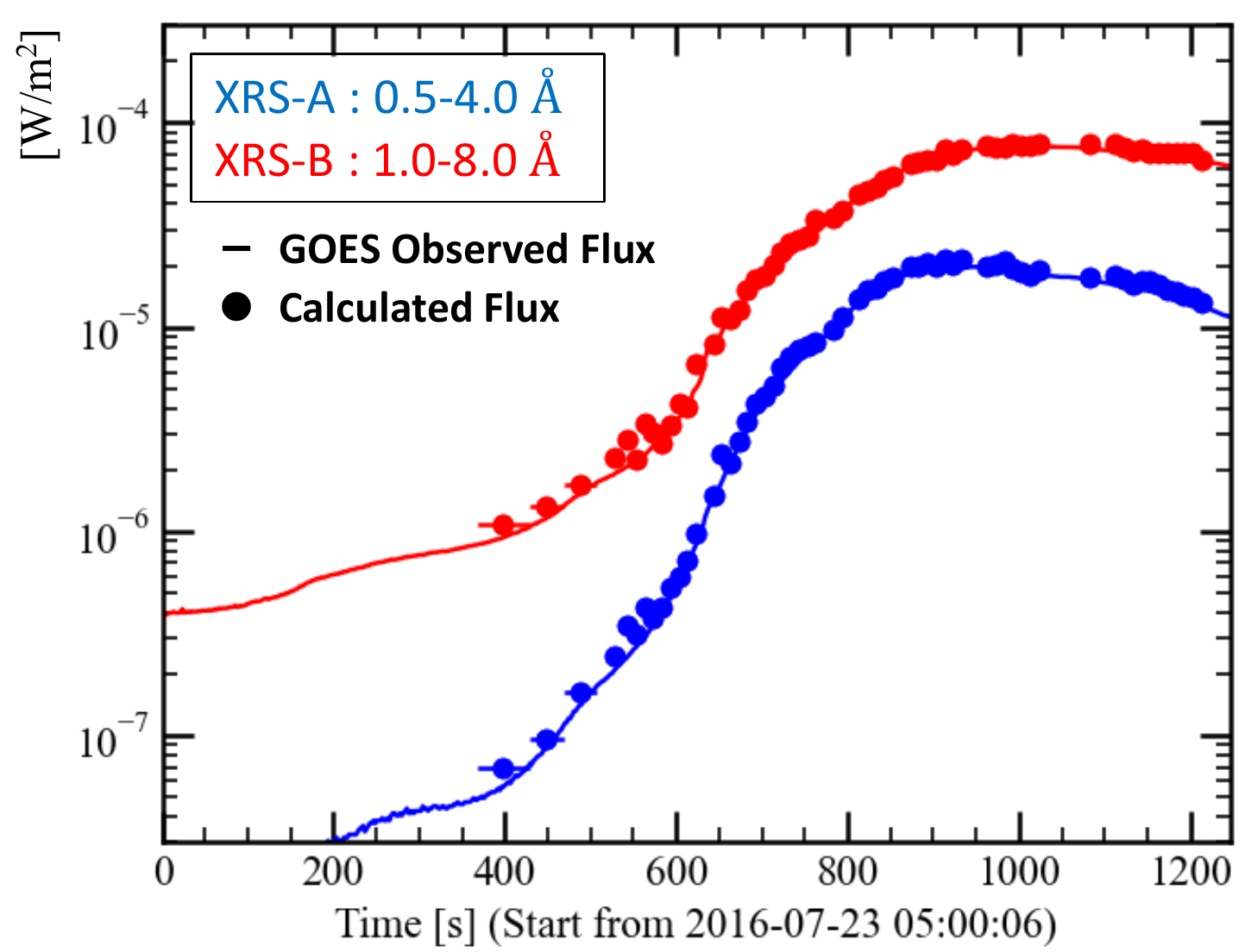}
\caption{Comparison of fluxes estimated from the results of spectral analysis to those actually observed by {\it GOES}. The model-estimated fluxes are in good agreement with the {\it GOES} observations.}
\label{goes_flux}
\end{center}
\end{figure}

Figure~\ref{goes_flux} shows the estimated {\it GOES} fluxes from the results of spectral analysis and the actually observed {\it GOES} fluxes. The estimated fluxes are consistent with the {\it GOES} observations throughout the flare. It should be noted that {\it MinXSS} and {\it RHESSI} data represent qualitatively different information compared with {\it GOES}. For {\it GOES}, with two broad channels, it is only possible to calculate the time evolution of the temperature and emission measure based on the observed fluxes in Figure~\ref{goes_flux} under the assumption of isothermal emission, and abundances must be assumed, typically as coronal \citep{white2005updated}. Therefore, this single temperature and emission measure just represents the averaged behavior of the thermal emission, like the dashed black curves in rows (a) and (b) in Figure~\ref{total_para}. In contrast, by using {\it MinXSS} and {\it RHESSI} data, we can resolve the multi-temperature structure of the thermal emission, as well as the non-thermal emission, and we can follow the time evolution of each component individually, like the colored measurements in Figure~\ref{total_para}.

\subsection{Microwave spectral analysis}\label{radio_spect_sec}
Microwave emission is observed during the flare by {\it NoRP} at frequencies from 1~GHz to 34~GHz, and we analyze its spectral evolution. We fit the NoRP spectra at frequencies of 2, 3.75, 9.4 and 17~GHz, with an integration time of 20~s, with a generic model function (see equation~(\ref{norp_model} in Appendix~\ref{app_radio}) during 05:08:49--05:14:49~UT, and determined the spectral index $\alpha_H$ above a turnover frequency $\widehat{\nu_T}$ for each time interval; the fit results for the spectral index are summarized in row (d) of Figure~\ref{total_para}.
During the fitting time interval, the turnover frequency $\widehat{\nu_T}$ is less than 17~GHz, and a negative spectral index ($-2.8 < \alpha_H <-1.4$) is determined at higher (optically-thin) frequencies. 

We also estimated the contribution of bremsstrahlung emission based on the temperatures and emission measures obtained by the X-ray spectral analysis using an optically-thin regime for hot and super-hot plasma and an optically-thick regime for cool plasma. The area of the cool plasma emission region estimated from the AIA 335~{\AA} image (95\% contour regions) are $A \approx 100~\mathrm{Mm^2}$. Therefore, the density of the cool plasma is estimated $ n_e \approx EM/A^{3/2} \approx 10^{12}~\mathrm{cm^{-3}}$, which is dense enough to be optically-thick ($\tau \approx 100$ for 17~GHz).
In the impulsive phase, the observed flux by {\it NoRP} at 17~GHz is $\sim$140~SFU, while the contribution of the bremsstrahlung emission is estimated to be less than 30~SFU and can be negligible.

Therefore, the continuous microwave emissions observed at 17~GHz must be optically-thin gyro-synchrotron emission from mildly relativistic non-thermal electron \citep{Dulk1985,bastian1998radio}.

\subsection{Imaging and DEM analysis}
\begin{figure*}[!hbt]
\begin{center}
\includegraphics[width=0.95\hsize]{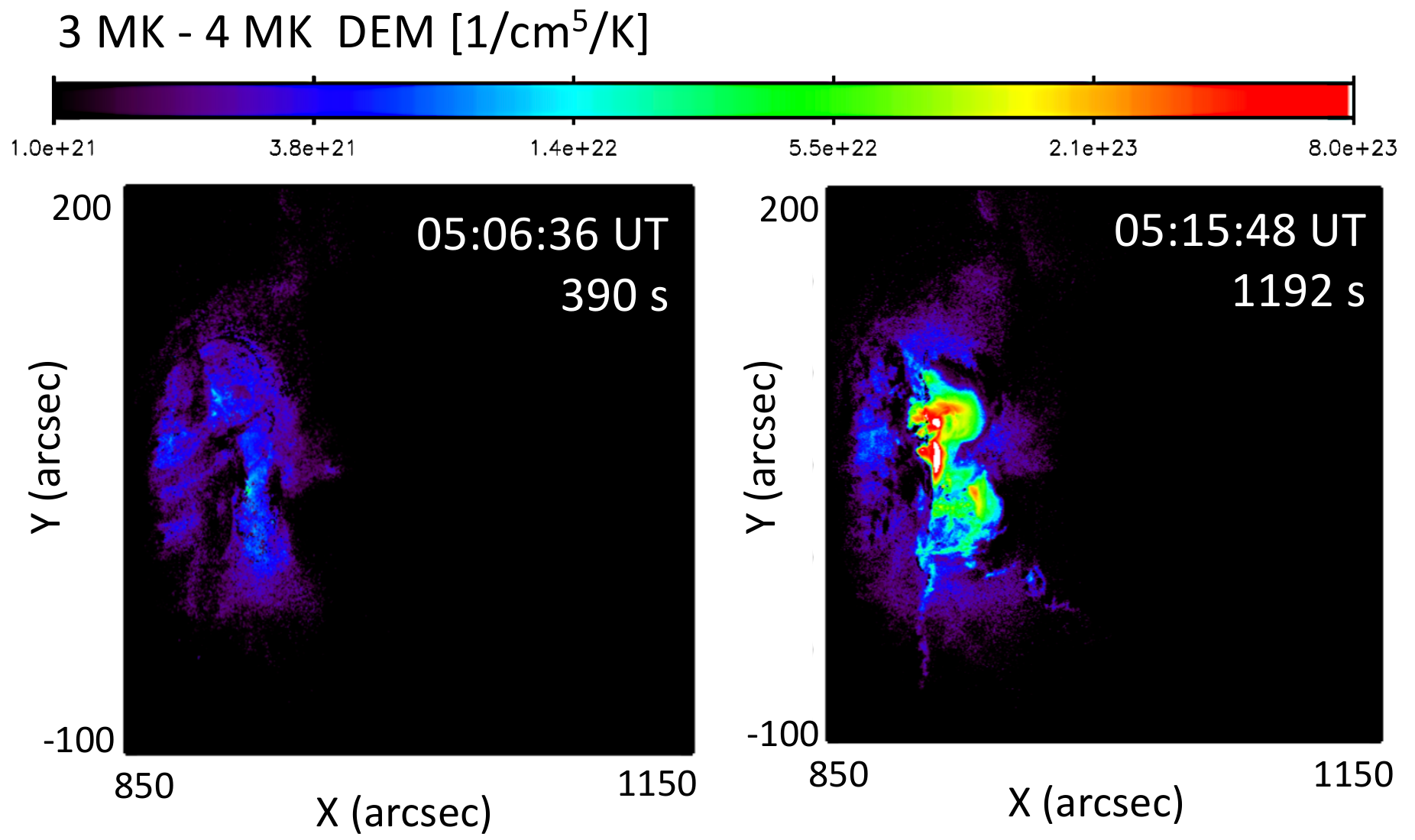}
\caption{Results of the AIA DEM calculation using the regularization method of \citet{hannah2012differential} at the beginning (05:06:36~UT) and end (05:15:48~UT) of the flare. Here, we use the six EUV filters with peak temperature sensitivity above 1~MK (94, 131, 171, 193, 211, and 333~{\AA}). Only the 3--4~MK temperature bin is shown. The DEM of 3--4~MK plasma clearly increases within the flaring loop. 
}
\label{dem}
\end{center}
\end{figure*}

To explore the locations of the thermal and non-thermal emission, we conducted an imaging analysis.
Using the six EUV filtergram observations of AIA with peak temperature sensitivity above 1~MK (94, 131, 171, 193, 211 and 335~{\AA}, corresponding to Fe lines from different ion species; \citealt{boerner2014}), the temperature distribution or ``differential emission measure'' (DEM) can be calculated by solving the relationship between the temperature response $K_i (T)$ in the $i$-th filter and the count rate $y_i$ observed in the $i$-th filter of AIA: 
\begin{equation}
y_{i} = \int_T K_i(T) DEM(T)dT
\end{equation}
Here, $DEM (T)$ is the Differential Emission Measure integrated along the line-of-sight to the observer:
\begin{equation}
DEM(T)dT = \int n^2_e(T)dz \hspace{10pt} \mathrm{[cm^{-5}]}
\end{equation}
where $n_e (T)$ is the thermal electron number density at temperature $T$.
In this study, we used the regularization method of \citet{hannah2012differential} to calculate the DEM in each pixel to identify the locations of cool thermal emission ($T = 3-4$~MK). Figure~\ref{dem} shows the results of the 3--4~MK temperature bin of the DEM calculation at the beginning of flare (05:06:36~UT) and at the end of flare (05:15:48~UT). The time evolution clearly shows that the 3--4~MK plasma increases in the flaring loop.

We note that, while AIA data can be used for DEM analysis in this way, the AIA response has poor sensitivity above $\sim$10~MK, particularly for the high flare temperatures observed here, and we therefore used AIA primarily to locate the cooler emission where the AIA response is strong. While AIA data could potentially be combined with the joint {\it MinXSS-RHESSI} data to further enhance the joint-instrument DEM analysis \citep[e.g.,][]{caspi2014b,inglis2014,moorephd}, such techniques are significantly complex and require careful consideration of the limitations of each instrument, and are thus beyond the scope of this work.

\begin{figure*}[!htb]
\begin{center}
\includegraphics[width=1.0\hsize]{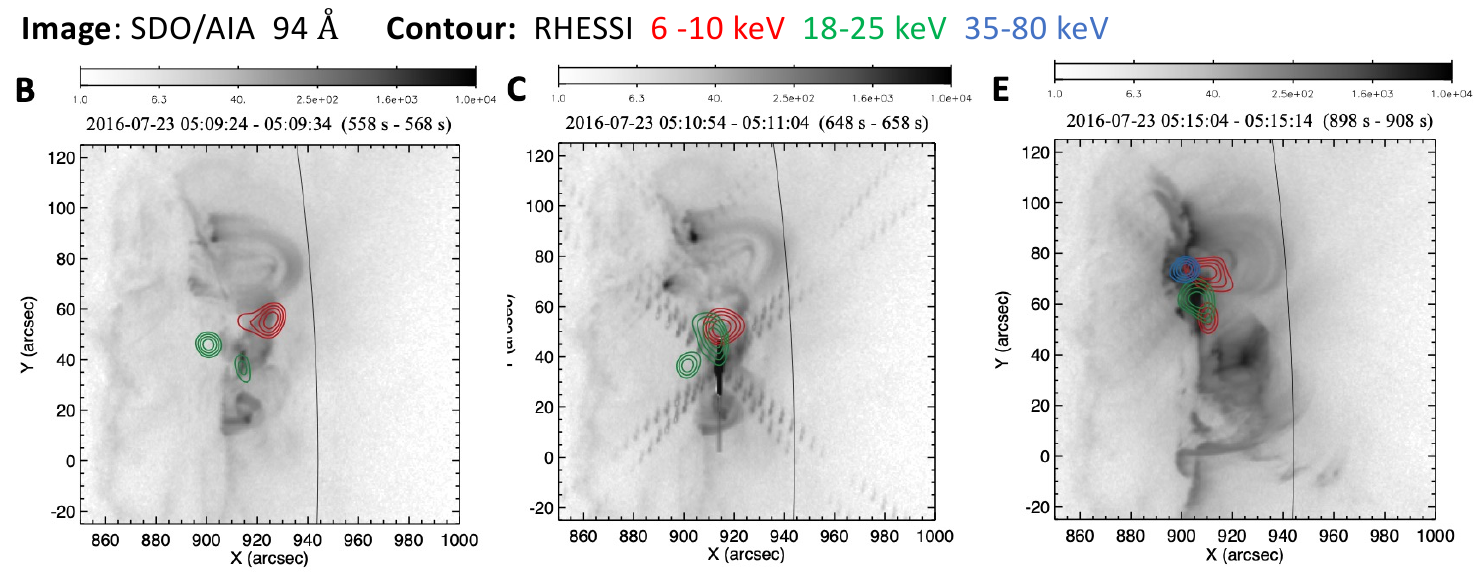}
\caption{{\it SDO}/AIA 94~{\AA} image (grayscale) overlaid with {\it RHESSI} HXR contours at 60\%, 70\%, 80\%, and 90\% of the peak intensity in the energy ranges of 6--10~keV (red), 18--25~keV (green), and 35--80~keV (blue).
The AIA 94~{\AA} emission corresponds primarily to plasma of $\sim$6~MK. The B, C, and E labels correspond to the time intervals shown in Figure~\ref{total_para}. {\it RHESSI} image synthesis was performed using the ``Clean" method \citep{hurford2003rhessi} using grids 3 and 8.
Image synthesis at 35--80~keV is omitted in panels B and C due to poor statistics at those times.}
\label{aia_image}
\end{center}
\end{figure*}

{\it RHESSI} uses rotation-modulation collimator to obtain spatial information, and the location of the X-ray emission on the Sun is calculated using an image synthesis method from flux modulations observed as the spacecraft rotates.
In order to locate the hot ($T \approx 15$~MK) and the super-hot ($T \approx 30$~MK) thermal components and the non-thermal component, {\it RHESSI} image synthesis was performed using the ``Clean" method \citep{hurford2003rhessi} available in SSW using imaging grids 3 and 8. The integration time was 40~s, and the energy bands were 6--10~keV, 18--25~keV and 35--80~keV. The 6--10~keV band is dominated by the hot thermal component, the 18--25~keV band has contributions from the non-thermal and (when present) the super-hot component, and the 35--80 keV band shows the non-thermal component only. The AIA 94~{\AA} images (emitted by \ion{Fe}{18}, corresponding to the plasma temperatures of 6~MK) with {\it RHESSI} contours overplotted are presented in Figure~\ref{aia_image}. The B, C, and E labels correspond to the time intervals shown in Figure~\ref{total_para}.

At the beginning of the flare (time interval~B), we can see the two footpoint HXR sources (18--25~keV) that correspond to the non-thermal thick-target bremsstrahlung when the accelerated electrons impact the chromosphere. The peak of the HXR emission (time interval~C) shows a dramatic increase in the HXR footpoint emission (18--25~keV) and most of the hot thermal emission (6--10~keV) is now lower in the loop. After the HXR peak (time interval~E), the emission region shifts from the south to the north, with an apparently different set of loops being energized and an additional source of hot emission (6--10~keV) appearing along with a new HXR footpoint (35--80~keV) and the super-hot thermal component (18--25~keV). We note that microwave imaging using {\it the Nobeyama Radioheliograph} \citep[{\it NoRH};][]{nakajima1994nobeyama} is not shown because the solar disk used for alignment and the flare microwave source could not be scaled simultaneously, and it is thus difficult to reliably compose the microwave image for this event.

\section{Discussion}\label{sec:Discussion}
By conducting simultaneous fitting using {\it MinXSS} and {\it RHESSI} spectra observed in the M7.6 flare that occurred on 2016 July 23, it becomes possible to clearly resolve a non-thermal power-law component and multiple thermal components (a cool plasma at $T \approx 3$~MK, a hot plasma  at $T \approx 15$~MK), and a super-hot plasma at $T \approx 30$~MK) and to follow their time evolution with a cadence of 10~s, which also corresponds to the Alfv\'en time scale in the solar corona. From the beginning of the flare, both the cool and hot thermal components are required to explain the observed spectra -- a single isothermal is not sufficient. As the non-thermal spectrum increases and hardens, the emission measures of both the hot and cool thermal components drastically increase, and images show that the cool plasma (3--4~MK) is confined to and increasing within the flaring loop. After that, the non-thermal emission softens, and the super-hot thermal emission ($T \approx 30$~MK) gradually increases, while continuous microwave emission -- optically-thin gyro-synchrotron emission from mildly relativistic non-thermal electrons -- is observed simultaneously. Subsequently, the HXR flux peaks a second time and the non-thermal emission hardens to its minimum spectral index of $\gamma_2 \approx 2.6$. 
Finally, as the non-thermal emission fades, each thermal emission gradually cools. The spectral analysis using {\it MinXSS} and {\it RHESSI} reproduces well the observed {\it GOES} SXR fluxes, providing confidence in the fit results.

This detailed time evolution information is a key to understanding the origins each spectral component. In particular, the emission measure of the cool thermal component drastically increases by more than two orders of magnitude in $\sim$300~s, and the temperature appears to decrease during the first HXR peak from Figure~\ref{total_para}.
The presence of cool thermal emission in solar flares was also noted by \citet{dennis2015solar}. From their spectrum obtained by SAX on {\it MESSENGER} during the 2007 June 1 M2.8 class flare, in the energy range of 1.5--8.5~keV, they reported that the spectrum was well described by both hot and cool thermal components and observed a similar drastic increase in emission measure and slight decrease in temperature for their cool plasma.
However, their results were obtained from spectra with a cadence of about 5~minutes. In contrast, through simultaneous observations and analyses of {\it MinXSS} and {\it RHESSI} spectra, we can track temperatures and emission measures with much higher cadence, every 10~s, for both the hot and cool thermal components. 

EUV imaging analysis shows that the DEM of 3--4~MK plasma, corresponding to the cool component in our spectral analysis, increases within the flaring loop (Figure~\ref{dem}).
This implies that the cool thermal component corresponds to plasma that fills the flaring loop associated with chromospheric evaporation. 
Similarly, the emission measure of the hot thermal component also drastically increases, by two orders of magnitude in $\sim$300~s, and is therefore also likely due to chromospheric evaporation.
Many prior analyses of {\it GOES} fluxes, {\it RHESSI} HXR, and EUV doppler-shift observations \citep[e.g.,][for reviews]{holman2011implications} have discussed chromospheric evaporation for hot plasma ($T = 10$--20~MK). 
However, by conducting SXR and HXR spectra analysis simultaneously, it is possible to reveal the multi-thermal structure of chromospheric evaporation including the cool plasma ($T \approx 3~\mathrm{MK}$) together with the hotter components, self-consistently. 
\citet{yokoyama1998two} conducted MHD simulations and suggested the presence of a cool plasma ($T = 3$--6~MK) for the large-scale flare events such as the arcade reformation associated with the prominence eruption. \citet{milligan2009velocity} also found blueshifts of \ion{Fe}{14}--\ion{}{24} line emission ($\sim$2--16~MK) in a flare, which is indicative of evaporated material at these temperatures. These studies are consistent with our evaporative interpretation of the cool thermal component.

The relationships between the evolution of the super-hot plasma and the microwave and hard X-ray emissions are keys to elucidate the potential origins of the super-hot plasma.
During the interval phase (spectrum D), we found that the emission measure of the super-hot plasma increased simultaneously with the continuous gyro-synchrotron emission at microwave frequencies. 
Since gyro-synchrotron emission is radiated from mildly relativistic non-thermal electrons in the presence of a magnetic field, this correlation implies that the super-hot component may originate from thermalization of the non-thermal electrons trapped in the corona, within the flaring loop. Similarly, the hard X-ray emission, which implies chromospheric energy deposition by accelerated electrons, shows only small sub-peaks, consistent with the microwave-generating electrons being largely trapped and not precipitating significantly into the chromosphere at this time. Moreover, in the decay phase, as the microwave emissions decreases, the temperature of the super-hot plasma gradually cools, and the emission measure also decreases.
\citet{caspi2010rhessi} found that, for their flare, the hot and super-hot plasma were spatially distinct, separated by $\sim$11.7{\arcsec} based on {\it RHESSI} imaging, and suggested that significant heating of super-hot plasma occurs directly in the corona. A more detailed analysis by \citet{caspi2015} found that the super-hot emission was also cospatial with apparent non-thermal emission for some time, indicating a potential link between super-hot plasma and non-thermal electrons accelerated in the corona. Our microwave results support that interpretation. 

We note that \citet{cheung2019comprehensive} suggested that adiabatic compression and viscous dissipation create a super-hot plasma with temperatures exceeding 100~MK at a higher altitude than hot plasma loop-tops, based on MHD simulations. \citet{caspi2015} also found that the super-hot source was at higher altitudes that the hot plasma, although cooler ($\sim$40~MK) than suggested by \citet{cheung2019comprehensive}.
In this flare, however, the apparent super-hot emission (18--25~keV contours in panel~E of Figure~\ref{aia_image}) appears to be at similar or possibly even lower altitude than the hot emission (6--10~keV contours in that Figure), although this is somewhat complicated by the non-thermal contribution to the 18--25~keV contours and by the more complex flaring geometry, with multiple loop systems being energized. A ``thermal imaging'' analysis to more cleanly separate the emission sources by combining the spectral and spatial information \citep[as in][]{caspi2015} may yield further insight and will be the subject of a future paper.

Since the dynamic range of {\it RHESSI} imaging is only about $\sim$10, it is hard to distinguish X-ray sources that are $\sim$10 times weaker than the brightest points, e.g., to distinguish weak coronal emission in the presence of bright chromospheric emission. While our speculation on the origins of the super-hot plasma are consistent with observations and prior interpretation, truly confirming the origin of super-hot plasma requires imaging spectroscopic observation with significantly improved dynamic range in both the SXR and HXR ranges. Elemental abundances are also valuable tracers of plasma origins \citep[e.g.,][]{warren2014,laming2021}, but many of the relevant lines (e.g., \ion{Fe}{25}) have a broad temperature response and will therefore contain contributions from both hot and super-hot plasmas, which can be difficult to distinguish in spatially-integrated spectra. 
The {\it Focusing Optics X-ray Solar Imager} \citep[{\it FOXSI};][]{krucker2009focusing} sounding rocket experiment demonstrated solar spectroscopy from directly-focused HXR observations. The upcoming {\it FOXSI-4} launch \citep{2020AGUFMSH0480011G,Camilo2021} will be specifically timed during a solar flare, achieving a much better dynamic range of up to $\sim$100 times larger than {\it RHESSI} to provide simultaneous diagnostics of spatially-separated coronal and chromospheric emission. Future photon-counting imaging spectrometers coupled with high-resolution focusing optics, in both soft and hard X-ray bands such as from the satellite mission concept {\it PhoENiX} \citep[Physics of Energetic and Non-thermal plasmas in the X (= magnetic reconnection) region;][]{narukage2019satellite} will enable us to follow the spectral evolution from different regions. This would be even further improved with higher spectral resolution across a wide band, to provide abundance diagnostics from a variety of elements (e.g., Fe, Ca, Si, Mg, O, Ne, Ar) across a broad range of coronal temperatures, such as would be provided by instruments like the {\it Marshall Grazing Incidence X-ray Spectrometer} \citep[{\it MaGIXS};][]{kobayashi2018,athiray2019} and the {\it CubeSat Imaging X-ray Solar Spectrometer} \citep[{\it CubIXSS};][]{cubixss2021}, for additional constraints on plasma heating mechanisms.
With such observations, it will be possible to verify and/or improve the interpretations of the origins of each thermal component we obtained from this study by directly comparing the emissions and their time evolution from the flaring loop, including footpoints and looptop, and any other relevant sources. 

\section{Conclusion}\label{sec:Conclusion}
In summary, we have conducted a wide-band X-ray spectral analysis using combined soft X-ray spectra from {\it MinXSS} and hard X-ray spectra from {\it RHESSI}, simultaneously. 
This joint analysis revealed  a non-thermal component and three-temperature multi-thermal component -- ``cool" ($T \approx$~3 MK), ``hot" ($T \approx$~15 MK), and ``super-hot" ($T \approx$~30 MK) -- with most of these components present throughout the flare.
In addition, we followed the time evolution of the thermal and non-thermal emissions with 10~s cadence, which corresponds to the Alfv\'en time scale in the solar corona. 
The time evolution of the spectral components and an imaging DEM analysis suggest that the cool and hot thermal components both correspond to plasma filling the flaring loop associated with chromospheric evaporation. On the other hand, a correlation between the super-hot thermal time evolution and microwave emission from non-thermal electrons suggests that the super-hot component could be explained by thermalization of the non-thermal electrons trapped in the flaring loop. 
Following the time evolution of the multi-temperature structure of the spectra using {\it MinXSS} and {\it RHESSI} provides new insights into the possible origins of these thermal emissions. However, it is necessary to follow the time evolution of the spectra in each emission region, with improved dynamic range in each spatial region, to elucidate the origin of the super-hot thermal component and its relationship with non-thermal emission.
In the future, direct-focusing SXR and HXR observations such as from {\it FOXSI-4} and {\it PhoENiX} will allow us to observe a wide dynamic range and track spectral evolution region by region.
By accumulating such observations, we will be able to clarify the origin of each thermal and non-thermal emission component and their relationship, which will help us understand the particle acceleration and resolve the complex energy conversion system associated with magnetic reconnection.

\acknowledgments
The authors would like to thank Christopher S. Moore for providing the response matrix of the {\it MinXSS} silicon drift detector (X123-SDD). The authors also thank Lindsay Glesener and Julie Vievering for their helpful discussions. This work was supported by JSPS, Japan KAKENHI Grant Number 17K14314, 18H03724, 18H05457, 20H00153, 21KK0052, 22H00134 and 22J12583. This work was also supported by FoPM (WINGS Program) and JSR Fellowship, the University of Tokyo, Japan. This work was carried out by the joint research program of the Institute for Space-Earth Environmental Research, Nagoya University. AC and TNW were supported by NASA grant NNX17AI71G, and AC was also partially supported by NASA grant 80NSSC19K0287.

\appendix
\section{XSPEC format and Data Preparation}\label{app_xspec}
XSPEC \citep[version 12.11.0;][]{arnaud1996xspec} is the standard spectrum analysis tool in the field of high-energy astronomy. To perform fitting using XSPEC, the following FITS (Flexible Image Transport System) files are required:
\begin{enumerate}
    \item Spectrum PHA data File (PHA File):\\
    The observation data (detected counts and statistical/systematic errors in each channel of the detector) and auxiliary information required during the spectral analysis process (live-time, observation time, and the related background and response files, etc.).
    \item Redistribution Matrix File (RMF):\\
    Information on nominal energy range in each channel of the detector and a two-dimensional matrix (Redistribution Matrix). The redistribution matrix (also sometimes called a ``detector response matrix'') represents the probability that an incident photon with energy $E$ will be detected in a certain channel $i$, accounting for potential energy loss or gain effects due to energy resolution, escape peaks, Compton scattering, etc. 
    \item Ancillary Response File (ARF):\\
    One-dimensional vector as a function of energy describing the effective area, such as filter transmission and detection efficiency of the detector.
\end{enumerate}

All of the above files must be created according to the OGIP standard data format described in \citet{arnaud2009ogip} and \citet{george2007calibration}. For solar observations, detector response information corresponding to the RMF and ARF is often already convolved into one two-dimensional data file and can be loaded as a single ``Response File" in XSPEC.

We created the PHA files and response files based on {\it MinXSS} and {\it RHESSI} data through the following procedure.
For {\it MinXSS}, the spectral data can be directly downloaded from the {\it MinXSS} website.
Various levels of data are pre-prepared in IDL `sav' file format, depending on the desired stage of data processing.  In this analysis, we used ``Level 0D'' data\footnote{\url{https://lasp.colorado.edu/home/minxss/data/level-0d}}, which includes raw detector count data along with the ancillary information (measurement time, satellite position, etc.) required for calibration and processing to higher levels. The spectrum, observation time, and integrated livetime information was extracted from the IDL sav file and the FITS files were created according to the OGIP data format.
For {\it RHESSI}, a dedicated software package is available within the IDL-based integrated solar analysis software SolarSoftWare (SSW) distribution, providing complete access to the data and all processing tools. The observation time, energy range and energy binning were specified in calls to the software to generate appropriate spectrum and response data. The FITS files produced by the {\it RHESSI} SSW package are designed to be compatible with OSPEX, the standard spectral analysis tool in SSW, and this format is not directly compatible with XSPEC. Therefore, we converted these FITS files to the OGIP data format for ingestion into XSPEC.

\section{Comparison among the multi-temperature model fitting}\label{app_spect}
\begin{figure*}[hbt]
\begin{center}
\includegraphics[width=1.0\hsize]{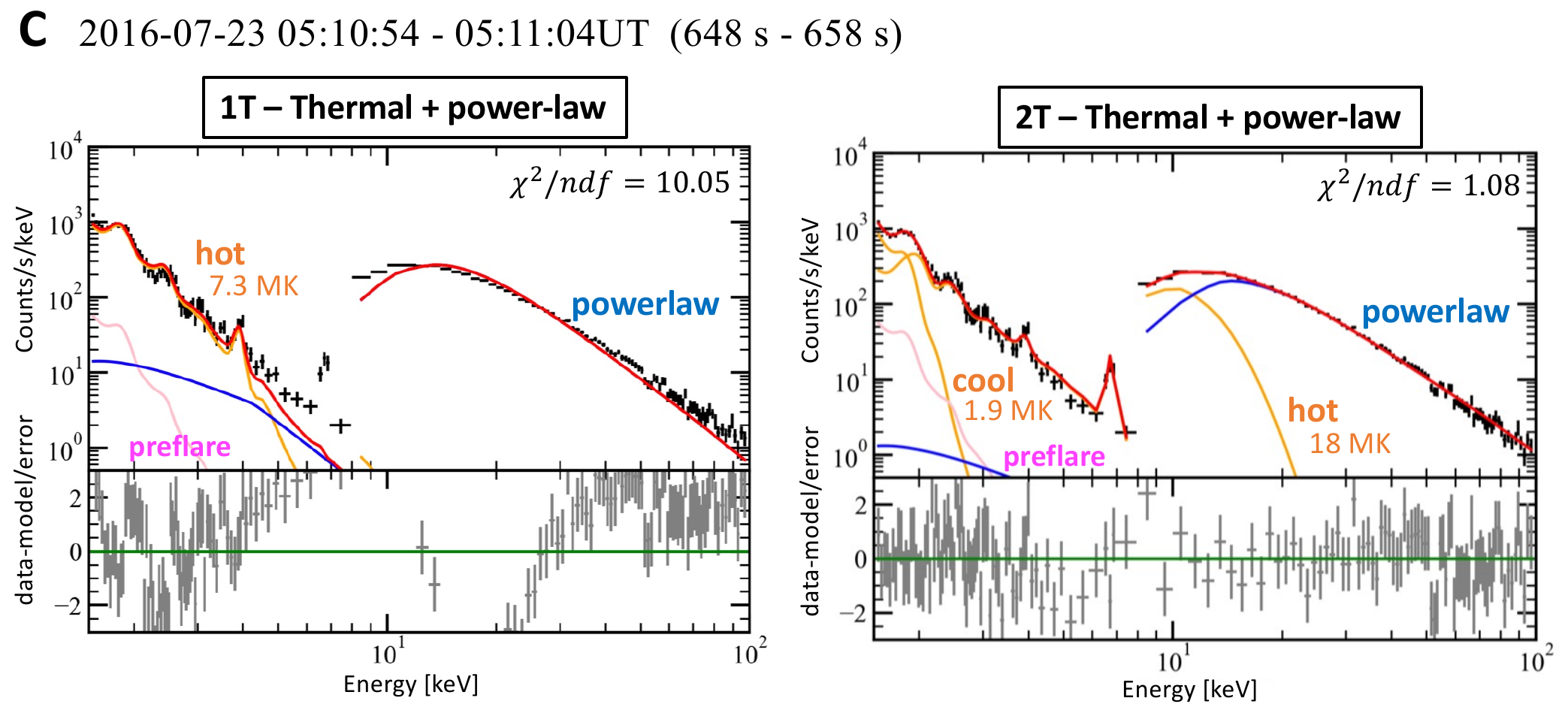}
\caption{{\it MinXSS} and {\it RHESSI} count flux spectra and two model fits for the 2016-07-23 05:10:54--05:11:04~UT period (time interval~C in Figure~\ref{total_para}).
Left: isothermal component with non-thermal power-law component and Right: two-temperature thermal components with non-thermal power-law component. Both models are fit over the entire 1.5--100 keV range. 
}
\label{2t_spect}
\end{center}
\end{figure*}

Figure~\ref{2t_spect} shows the ``isothermal" and ``two-temperature" thermal emission models with non-thermal power-law model fits to the observed {\it MinXSS} and {\it RHESSI} count flux spectra for the 2016-07-23 05:10:54--05:11:04~UT period (time interval~C in Figure~\ref{total_para}). The fitting is performed over the 1.5--100~keV energy range and the normalized residuals (the differences between the observed count flux and best-fit model count flux, divided by the statistical 1$\sigma$ uncertainties) are shown below. For the isothermal model, a good fit could not be achieved, especially over the 1.5--20~keV SXR energy range. An additional ``cool" thermal component ($T \approx 3$~MK) is required to explain the observed spectrum, as shown in the ``two-temperature'' model fit. Therefore, in this analysis, the ``cool" thermal emission component was always included in the fitting model during the time range of 05:06:14--05:29:24~UT, when spectral analysis is conducted using both {\it MinXSS} and {\it RHESSI}.

\begin{figure*}[hbt]
\begin{center}
\includegraphics[width=1.0\hsize]{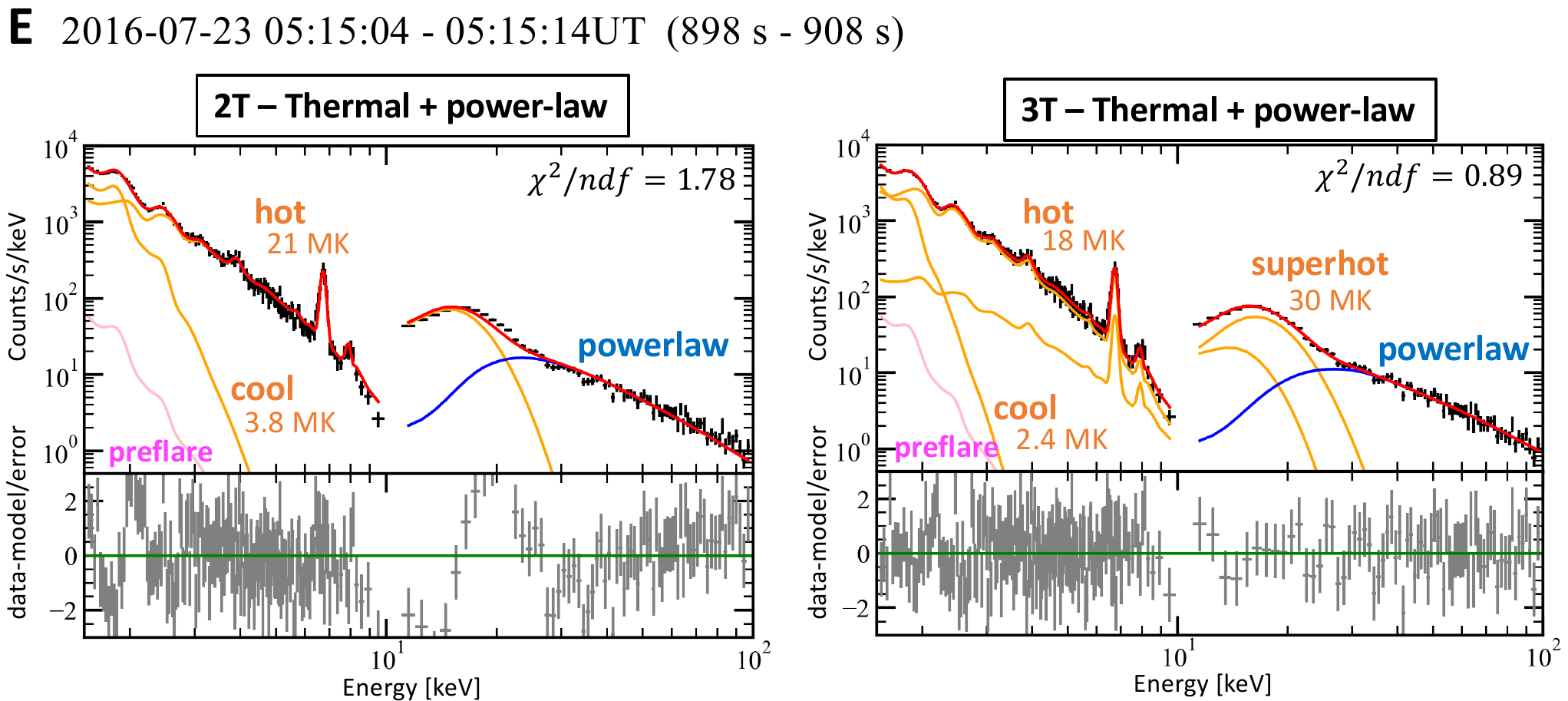}
\caption{{\it MinXSS} and {\it RHESSI} count flux spectra and two model fits for the 2016-07-23 05:15:04--05:15:14~UT period (time interval~E in Figure~\ref{total_para}).
Left: two-temperature thermal components with non-thermal power-law component and Right: three-temperature thermal emission with non-thermal power-law model. Both models are fit over the entire 1.5--100 keV range. 
}
\label{3t_spect}
\end{center}
\end{figure*}

Figure~\ref{3t_spect} shows the ``two-temperature" and ``three-temperature" thermal emission models with non-thermal power-law model fits to the observed {\it MinXSS} and {\it RHESSI} count flux spectra for the 2016-07-23 05:15:04--05:15:14~UT period, near the peak of the flare (time interval~E in Figure~\ref{total_para}). During the flare peak, the ``cool" and ``hot" thermal components are not sufficient to explain the observed spectrum, particularly from 10 to 30~keV, and an additional ``super-hot" thermal component ($T \approx 30$~MK) is required to improve the fit, as seen both in the overall reduced $\chi^2$ statistic as well as the behavior of the normalized residuals. Therefore, in this analysis, the three-temperature thermal model is used for times after 05:11:14~UT.

\section{Microwave spectral fitting}\label{app_radio}
\begin{figure}[htb]
\begin{center}
\includegraphics[width=1.0\hsize]{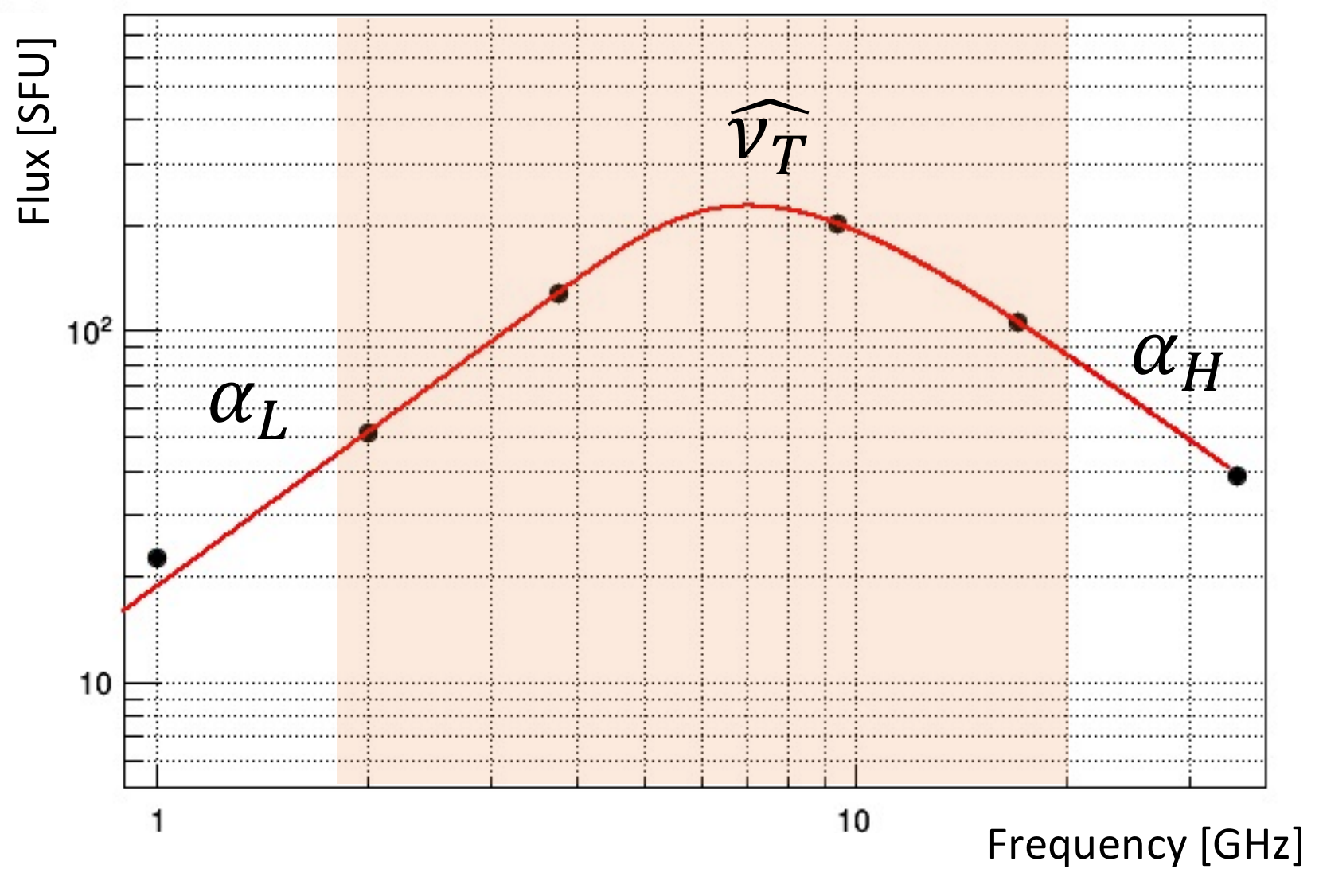}
\caption{Microwave spectrum taken with {\it NoRP} during
the 2016 July 23 M7.6 solar flare at 05:12:19--05:12:39~UT. The microwave spectra are integrated over 20~s to improve statistics and are fit using the model described by equation~(\ref{norp_model}) over the 2--17~GHz range.}
\label{norp_spect}
\end{center}
\end{figure}

In this study, we fit the NoRP spectra at frequencies of 2, 3.75, 9.4 and 17~GHz, with integration times of 20~s, using a generic model function \citep{stahli1989high,silva2000correlation}:
\begin{equation}
F_{v} = N\left(\frac{\nu}{\widehat{\nu_{T}}}\right)^{\alpha_L}\left[1-\exp\left[-\left(\frac{\nu}{\widehat{\nu_{T}}}\right)^{\alpha_H-\alpha_L}\right]\right] \approx \begin{cases}
N(\nu/\widehat{\nu_T})^{\alpha_L}, & {\rm for} ~~\nu \ll \widehat{\nu_T} \\
N(\nu/\widehat{\nu_T})^{\alpha_H} & {\rm for} ~~ \nu \gg \widehat{\nu_T} 
\end{cases}
\label{norp_model}
\end{equation}
where $\nu$ is the frequency and $\widehat{\nu_T}$ is a turnover frequency, and $\alpha_L$ and $\alpha_H$ are the spectral indices at frequencies lower (optically-thick) and higher (optically-thin), respectively, than the turnover. Figure~\ref{norp_spect} shows the fitted spectrum at 05:12:19--05:12:39~UT, as an example. 

\bibliographystyle{aasjournal}
\bibliography{main}

\end{document}